\documentclass[aps,prx,reprint,floatfix,superscriptaddress]{revtex4-2}
\usepackage{graphicx}
\usepackage{hyperref}
\usepackage{color}
\usepackage{amsmath}
\usepackage{amssymb}
\usepackage{braket}
\usepackage{ulem}
\usepackage{xcolor}
\usepackage{lineno}

\begin{document}

\title{Scalable Parameter Design for Superconducting Quantum Circuits with Graph Neural Networks}

\author{Hao Ai}
\affiliation{School of Integrated Circuits, Tsinghua University, Beijing 100084, China.}
\author{Yu-xi \surname{Liu}}
\altaffiliation[yuxiliu@mail.tsinghua.edu.cn]{}
\affiliation{School of Integrated Circuits, Tsinghua University, Beijing 100084, China.}
\date{\today}

\begin{abstract}
To demonstrate supremacy of quantum computing, increasingly large-scale superconducting quantum computing chips are being designed and fabricated.
However, the complexity of simulating quantum systems poses a significant challenge to computer-aided design of quantum chips, especially for large-scale chips. 
Harnessing the scalability of graph neural networks (GNNs), we here propose a parameter designing algorithm for large-scale superconducting quantum circuits. The algorithm depends on the so-called 'three-stair scaling' mechanism, which comprises two neural-network models: an evaluator supervisedly trained on small-scale circuits for applying to medium-scale circuits, and a designer unsupervisedly trained on medium-scale circuits for applying to large-scale ones. 
We demonstrate our algorithm in mitigating quantum crosstalk errors. Frequencies for both single- and two-qubit gates (corresponding to the parameters of nodes and edges) are considered simultaneously. 
Numerical results indicate that the well-trained designer achieves notable advantages in efficiency, effectiveness, and scalability. 
For example, for large-scale superconducting quantum circuits consisting of around $870$ qubits, our GNNs-based algorithm achieves $51\%$ of the errors produced by the state-of-the-art algorithm, with a time reduction from $90\ {\rm min}$ to $27\ {\rm sec}$. 
Overall, a better-performing and more scalable algorithm for designing parameters of superconducting quantum chips is proposed, which demonstrates the advantages of applying GNNs in superconducting quantum chips.
\end{abstract}

\maketitle

\textit{Introduction--}Quantum computing may have applications in cryptography~\cite{IntroQuUse_ShorAlg}, databases~\cite{IntroQuUse_GroverAlg}, computational chemistry~\cite{IntroQuUse_Chemistry}, and machine learning~\cite{IntroQuUse_QML}. Thus, large-scale processors are demanded to manifest the quantum advantages. Among various platforms including neutral atom arrays~\cite{IntroQuRealize_atom_Lukin2024} and trapped irons~\cite{IntroQuRealize_ion_1st2021}, superconducting quantum computing system is a leading candidate~\cite{IntroQuRealize_SQC_google2019,IntroQuRealize_SQC_zhuxiaobo2021,IntroQuRealize_SQC_IBM2023,IntroQuRealize_Gu2017,IntroQuRealize_zhang2023,IntroQuRealize_Xu2023}. The scales of quantum processors implemented by superconducting quantum electronic circuits (SQECs) are rapidly increasing. The numbers of qubits are now more than hundreds and even reach thousands, e.g. in chips of IBM company~\cite{IntroQuRealize_IBM1000_2023news}. Due to the rapid growth of circuit complexity, the manual design on these large-scale SQECs are becoming challenging. Thus, electronic design automation for quantum computing systems aided by electronic computers or artificial intelligence is attracting more and more attention~\cite{IntroAI4Q_QX_QCsimu,IntroAI4Q_ML_QC_prxq,IntroAI4Q_ML_QEC_SQC,IntroAI4Q_ML_QC_ieee,IntroAI4Q_QC_diffusion,IntroAI4Q_QControl,IntroAI4Q_RL,IntroAI4Q_RL_expSQC}.
However, most of the discussions are limited to the small or medium scales. Because the dimension of the Hilbert space exponentially scales with the number of qubits, the evaluation of the designed SQECs through numerical simulations is becoming a challenging task, which further complicates the automated design towards large-scale quantum processors.

Meanwhile, we notice some neural network models with scalable properties. For example, transformers can deal with texts of different lengths~\cite{scalable_transformer}, and diffusion models can generate images with different pixel numbers~\cite{scalable_diffusion}. Within the scalable neural networks, graph neural networks~\cite{GNN_GCN,GNN_GAT} (GNNs) have demonstrated advantages in a wide range of scientific and engineering applications~\cite{GNN_apply}. This inspires us to achieve automated design for large-scale SQECs by conceptualizing SQECs as graphs and leveraging the scalability of GNNs.

We here focus on the automated design of working parameters in SQECs with given structures.
The configuration of parameters on each qubit can significantly impact the performance of quantum computing. Especially some of the parameters may have graph-dependent effects and need to be configured differently on different qubits.
In previous studies, the Snake algorithm was developed for designing parameters in large-scale SQECs~\cite{Introsnake_1,Introsnake_2} and was applied in Google's Sycamore processor~\cite{IntroQuRealize_SQC_google2019}.
However, the Snake algorithm ignores the detailed graph structures of the SQECs and relies on multiple calls to optimizers~\cite{scipy}. These characteristics may impact the effectiveness and efficiency of the Snake algorithm. Additionally, some other algorithms are also proposed for the parameter designing task~\cite{Intro_except_snake1,Intro_except_snake2,Intro_except_snake3}. However, they do not demonstrate their application to large-scale SQECs.

In this study, a GNNs-based algorithm for designing parameters in SQECs is proposed. The realization of the algorithm requires two neural-network models, an evaluator for estimating the quality of parameter assignments, and a designer for designing an optimal parameter assignment based on the evaluator's feedback.
The influence of considered parameters is assumed to be localized. Therefore, the evaluator, comprising some multi-layer perceptrons~\cite{MLP_1st,MLP_bp} (MLPs), is supervisedly trained on several small-scale SQECs, to avoid direct simulations on larger quantum processors.
Utilizing the errors computed by the evaluator, the GNNs-based designer is unsupervisedly trained on randomly generated medium-scale SQECs with around $100$ qubits. 
Leveraging the scalability of GNNs, the trained designer is directly applied to different graphs in terms of structure or scale, for example, the large-scale SQECs containing nearly $1000$ qubits.
The proposed algorithm is demonstrated through the task of designing frequency assignments to mitigate quantum crosstalk errors. As quantum crosstalk still exists in ideal systems, it is possible to model the errors using numerical simulations~\cite{Crosstalk_zhaopeng} instead of real experiments. 
According to our numerical results, the well-trained GNNs-based designer demonstrates significant advantages over previous methods in terms of both efficiency and effectiveness for all tested graph scales, ranging from $31$ to $870$ qubits. For instance, in large-scale SQECs comprising $870$ qubits, the designer achieves an average crosstalk error of $0.0090$ within $27$ seconds, whereas the Snake algorithm takes $90$ minutes to reach an average error of $0.0178$.

The contributions of the paper mainly include:
(1) To the best of our knowledge, GNNs are applied to the hardware-level design of quantum chips for the first time, demonstrating the advantages in performance and scalability.
(2) A graph-based scalable evaluator is trained, which can estimate quantum crosstalk errors for large-scale SQECs. While the previous numerical discussions are only limited to small-scale~\cite{Crosstalk_zhaopeng}. 
(3) A GNNs-based designer is trained to assign parameters for SQECs, showcasing the advantages in effectiveness, efficiency, and scalability over alternative methods~\cite{Introsnake_2,Intro_except_snake1,Intro_except_snake2,Intro_except_snake3}.
Note that the focused parameter designing task is a necessary process when utilizing a quantum chip. 

\begin{figure}[tbp]
	\centering
	\includegraphics[width=1\linewidth]{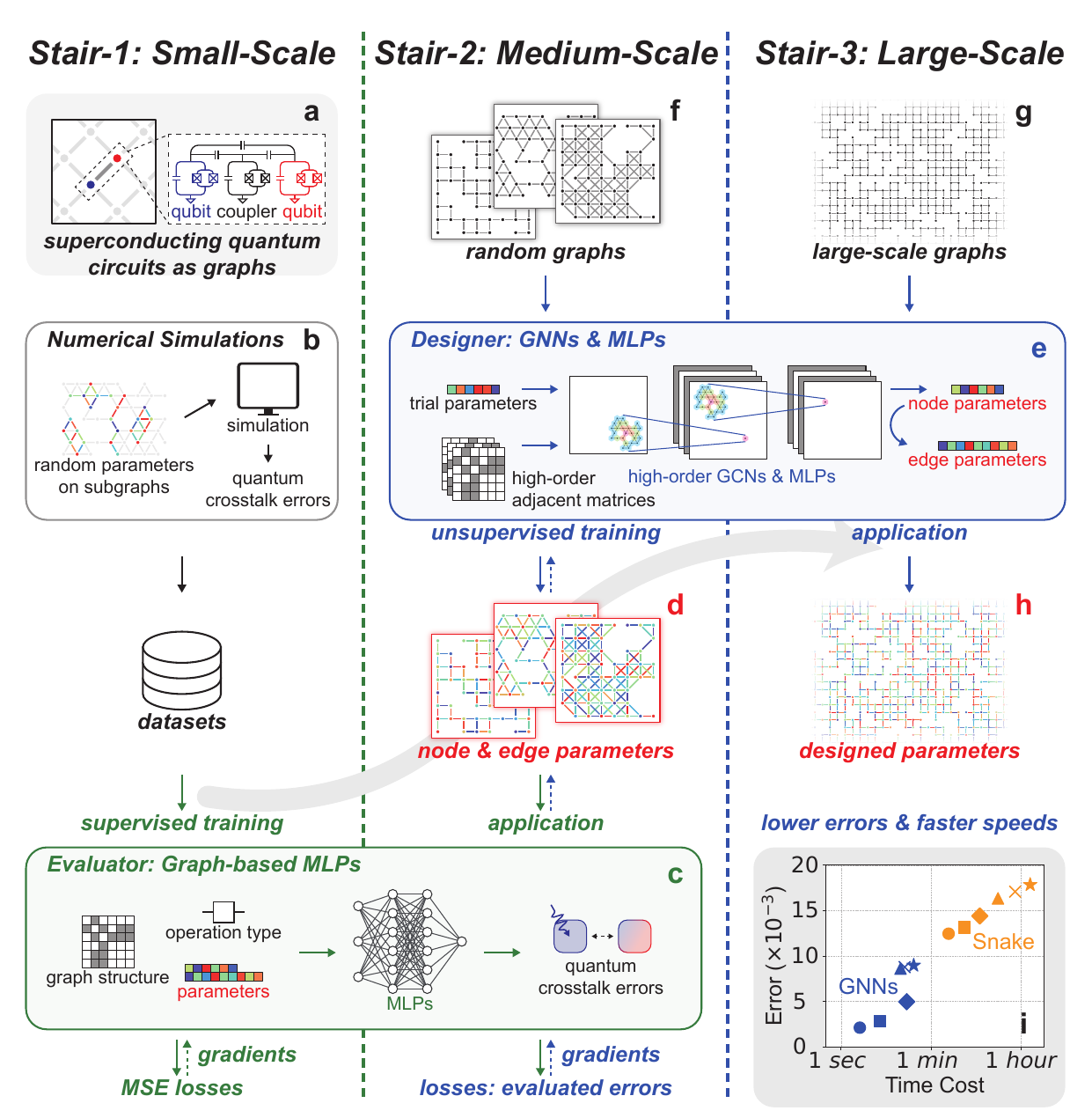}\\
	\caption{
		In the 'three-stair scaling' technique, two neural-network models, i.e., the evaluator and the designer, are trained and applied to connect the three scales.
		The graphic representation of SQECs is shown in \textbf{(a)}, where graph's nodes and edges are represented as dots and segments.
		In \textbf{(b)}, random parameters are sampled on several different small-scale subgraphs (e.g., the colored parts, each with $6$ qubits) in a larger-scale graph.
		Quantum crosstalk errors are numerically solved to construct datasets for the evaluator.
		The evaluator, illustrated in \textbf{(c)}, utilizes multiple graph-based MLPs to predict quantum crosstalk errors.
		The MLPs are supervisedly trained according to the MSE losses.
		The evaluator trained on small-scale graphs is applied to medium-scale graphs, e.g., \textbf{(d)} with around $54$ qubits.
		Then using the evaluated errors and their gradients, the designer shown in \textbf{(e)} is unsupervisedly trained on random medium-scale graphs such as the graphs in \textbf{(f)}.
		The designer assigns parameters for \textbf{(f)} to obtain \textbf{(d)}, which integrates GNNs and MLPs to determine optimal node and edge parameters based on the input graphs and trial parameters.
		Finally, the trained designer can be directly applied to graphs of various structures and scales, including large-scale ones demonstrated with \textbf{(g)} and \textbf{(h)}, e.g., a SQEC with $870$ qubits.
		The comparison of our algorithm (GNNs, with blue markers) and traditional algorithm (Snake, with orange markers) is plotted in \textbf{(i)}.
		The tested graphs range from medium-scale to large-scale, and different scales are represented by different markers, e.g., $\blacktriangle$ and $\bigstar$.
		More details about the tests can be found in Fig.~\ref{fig2}, and the well-trained designer achieves notable advantages in both efficiency and effectiveness.
		}\label{fig1}
\end{figure}


\textit{Algorithm--}Our GNNs-based algorithm can design parameters for a coupled-qubit graph conceptualized from a given SQEC, as shown in Fig.~\ref{fig1}(a). Qubits are composed of several circuit elements and serve as the fundamental units for storing and processing quantum information~\cite{transmon}. Different qubits are coupled through couplers and capacitors, enabling interactions necessary for quantum computation~\cite{coupler}. In the mapping from SQECs to graphs, each node represents one qubit and each edge corresponds to a pair of directly-coupled qubits. Both node and edge parameters need to be designed for mitigating errors of quantum computing~\cite{Introsnake_2,Crosstalk_zhaopeng}.
Some of the parameters are graph-independent, which can be optimized individually. While other parameters may have graph-dependent effects, i.e., the parameters of certain node or edge may impact the neighboring nodes and edges. This requires a comprehensive consideration of the entire graph during the designing process. Furthermore, the error dependence on the graph-dependent parameters is complicated, for example, parameters optimized for one node may cause large errors on other nodes. Consequently, powerful algorithms should be implemented to achieve the optimal design of the parameters.
Mathematically, the parameter designing task of SQECs can be interpreted as an extension of graph coloring problem~\cite{GCP_1st,GCP_amazon}, which involves node and edge colors, continuous coloring, and the impact of higher-order neighbors.

The proposed algorithm designs parameters for both nodes and edges, and considers errors from influences of higher-order neighbors (e.g., 2-hop, 3-hop neighbors).
As illustrated in Fig.~\ref{fig1}, graphs with three scales are involved, i.e., small-scale graphs denoted as stair-$1$, medium-scale graphs denoted as stair-$2$, and large-scale graphs denoted as stair-$3$. The three stairs are connected via two neural-network models: the evaluator and the designer. The evaluator is supervisedly trained on small-scale graphs, and is applied to medium-scale graphs. While the designer is unsupervisedly trained on medium-scale graphs to be applied to large-scale graphs.
First, as shown in Fig.~\ref{fig1}(b), training data for the evaluator is collected on several small graphs, where a variety of parameter assignments are uniformly sampled over an experimentally-available range as input labels, and corresponding errors are numerically simulated to serve as output labels.
Then the evaluator is trained through supervised learning.
As demonstrated in Fig.~\ref{fig1}(c), the evaluator takes the graph structure, parameter assignment, and type of quantum operations as inputs. It employs multiple graph-based MLPs to predict the errors from a source node (or edge) to a target node, which can be summed to obtain the total errors at every target node.
When training the evaluator, mean square errors (MSEs) are calculated as loss functions.
After training the evaluator, it can be directly applied on medium-scale graphs, e.g., Fig.~\ref{fig1}(d).
Unsupervised learning is employed to train the designer using the errors obtained by the evaluator.
As illustrated in Fig.~\ref{fig1}(e), the designer integrates GNNs and MLPs to determine optimal parameters for every node and edge based on the input graphs and trial parameters.
In each training iteration, a batch of medium-scale graphs is randomly created as input data for the designer, e.g., in Fig.~\ref{fig1}(f), where some of the nodes and edges are randomly removed from several fundamental graphs.
Subsequently, the designer outputs a parameter assignment for each graph as shown with Fig.~\ref{fig1}(d), and the corresponding average error is estimated using the trained evaluator.
After training the designer on medium-scale random graphs, the scalability of GNNs enables the trained designer to be directly applied to graphs of various structures and scales, including large-scale ones as exemplified with Figs.~\ref{fig1}(g) and (h). Additionally, leveraging the parallelism capabilities of  PyTorch~\cite{PyTorch} or TensorFlow~\cite{TensorFlow}, we can further enhance the performance on specific graphs by generating diverse sets of trial parameters and utilizing the designer in batches.
The demonstrated results of effectiveness and efficiency are shown in Fig.~\ref{fig1}(i).


\textit{Demonstration--}The algorithm is demonstrated with the task of designing frequencies in SQECs to mitigate quantum crosstalk errors. Node and edge frequencies are selected as parameters to be designed.
Quantum crosstalk is commonly present in SQECs~\cite{Crosstalk_zhaopeng,Crosstalk_learning,Crosstalk_freq,Crosstalk_2,Crosstalk_3,Crosstalk_4,Crosstalk_5,Crosstalk_6}.
It comes from some fundamental theories in quantum physics~\cite{SW_transfomation,dressed_states}, thus crosstalk errors are challenging to be completely eliminated, even in the ideal systems. 
Quantum crosstalk refers to the undesired interference experienced by the target qubits when quantum operations are applied to the source qubits.
Three types of quantum operations are involved in this study, including single-qubit $R_X$ (for nodes), single-qubit $R_Y$ (for nodes)~\cite{NielsenBook}, and two-qubit $R_{XY}$ via the $XY$ interaction (for edges)~\cite{iSWAP}. These operations can be used to build universal quantum computation.
The average crosstalk excitations from the source node $\mathcal{N}_i$ (or edge $\mathcal{E}_{ij}$) to the target node $\mathcal{N}_k$ solely depend on the operation type, graph structure, and frequency assignment~\cite{supply}. These dependencies define the inputs and architecture of the evaluator.
Extensive datasets are generated for training, validating, and testing the evaluator.
Various factors are considered in numerical simulations, such as residual couplings between high-order neighboring qubits~\cite{Crosstalk_zhaopeng,nnn_coupling}, the couplers connected between qubits~\cite{coupler}, and higher excitations of these quantum modes~\cite{transmon}.

\begin{table}[tbp]
	\caption{\label{table1} Results of the evaluator in the form of MSE($r^2$). }
	\begin{ruledtabular}
		\begin{tabular}{lccc}
			\textbf{Dataset}\footnote{Totally $15$ datasets are utilized, encompassing $5$ purposes and three types of quantum operations.}
			& \textbf{$R_X$} & \textbf{$R_Y$} & \textbf{$R_{XY}$} \\
			\colrule
			training            & 0.026(0.94) & 0.026(0.94) & 0.033(0.94) \\
			validation          & 0.027(0.94) & 0.029(0.94) & 0.030(0.95) \\
			testing             & 0.028(0.94) & 0.029(0.93) & 0.026(0.95) \\
			testing (NS)\footnote{Testing on \underline{N}ew \underline{S}tructures.}
						        & 0.034(0.92) & 0.024(0.94) & 0.041(0.94) \\
			testing (30Q)\footnote{Testing on a linear chain of \underline{$30$} coupled \underline{Q}ubits.}
								& 0.007(0.97) & 0.012(0.95) & 0.036(0.91) \\
		\end{tabular}
	\end{ruledtabular}
\end{table}


\textit{Results of the evaluator--}After generating the datasets, the evaluator is trained, validated, and tested in a supervised learning manner. The evaluator consists of multiple MLPs. Each MLP serves for one type of quantum operations and predicts the average crosstalk excitation on $\mathcal{N}_k$ resulting from operations on $\mathcal{N}_i$ (or $\mathcal{E}_{ij}$), where $k\neq i,j$. The inputs of the MLP are the relevant node and edge frequencies denoted by $\vec{\omega}$ and the distances represented by the elements of adjacency matrices, i.e., $A^p_{ik}$ (together with $A^p_{jk}$ for two-qubit operations).
To maintain consistency in the outputs of the MLPs, the crosstalk errors are transformed using base-$10$ logarithms. 
The losses are determined by the corresponding MSEs, $\mathcal{L}_O^{\rm evaluator} = |\Lambda_O|^{-1} \sum_{\lambda \in \Lambda_O} \delta_{O,\lambda}^2 $, and
\begin{equation}
	\delta_{O,\lambda}  = |{\rm MLP}_O(\vec{\omega}_\lambda, A^1_{\lambda,ik}, A^2_{\lambda,ik}, \cdots) - {\rm log}_{10} (E_{O,\lambda})|,
\end{equation}
where $O$ is the type of quantum operations, $\lambda$ is a specific index of data in the operation's dataset $\Lambda_O$ with the size $|\Lambda_O|$, and $E_{O,\lambda}$ is the simulated crosstalk error.
The crosstalk errors smaller than, e.g., $1\times10^{-4}$ are not concerned, because excessively small crosstalk errors usually do not dominate the error components.
While errors larger than, e.g., $1\times 10^{-1}$ are capped at $1\times 10^{-1}$ to avoid fitting out-of-distribution data.
Individual MLPs are supervisedly trained for different types of quantum operations, with the resulting MSEs and $r^2$ for the training, validation, and testing sets presented in Tab.~\ref{table1}. Additional tests on new graph structures (beyond the training graphs) and larger graph scales (than the small training graphs) ensure the evaluator's generalization ability~\cite{supply}. Each training set has $61,440$ data points, each testing and validation set on small graphs has $12,800$ data points, while each testing set on larger graphs has approximately $1,000$ data points. The coefficient of determination $r^2$ exceeds $0.9$ for each case, indicating a strong ability of the evaluator to fit the datasets~\cite{r2}. 

\begin{figure}[tbp]
	\centering
	\includegraphics[width=1\linewidth]{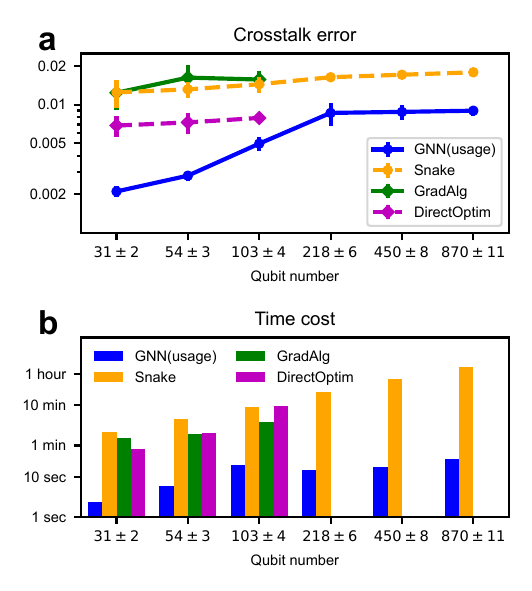}\\
	\caption{
		\textbf{Effectiveness, efficiency and scalability.}
		These are the detailed versions of Fig.~\ref{fig1}(i).
		\textbf{(a)} The average crosstalk errors using different methods on SQECs with various structures and scales.
		\textbf{(b)} The real-world running times of different algorithms.
		All the tests are performed using the same CPU and the same evaluator.
		Only the GNNs-based designer and the Snake algorithm are available for the large-scale graphs.
	}\label{fig2}
\end{figure}

\textit{Results of the designer--}As shown in Fig.~\ref{fig1}, during training the designer, medium-scale graphs $\mathcal{G} = (\mathcal{N}, \mathcal{E})$ with nodes $\mathcal{N}$ and edges $\mathcal{E}$ are randomly generated and input into the designer to output a frequency assignment $\vec\omega$ for each graph. The trained evaluator estimates the crosstalk errors for a particular $\vec\omega$ on corresponding $\mathcal{G}$. Then the designer's loss could be calculated,
\begin{equation}
\mathcal{L}^{\rm designer}(\mathcal{G},\vec{\omega}) = w_X \bar{E}_{R_X} + w_Y \bar{E}_{R_Y} + w_{XY} \bar{E}_{R_{XY}},
\end{equation}
where $\bar{E}_{O}$ is the average error for each target node under operation $O$~\cite{supply}. The default setting of the error weights $w_O$ is $w_X=w_Y=w_{XY}=1$.
The randomly-generated training graphs contain around $54$, $103$ and $218$ nodes, and the training batch size is $288$.
Approximately $700,000$ training steps were executed for about $2$ days with an RTX 4090 GPU. 
The trained designer can be readily applied to graphs of different structures and scales. 
We tested totally $30\times6$ random graphs with average qubit numbers of $32$, $54$, $103$, $218$, $450$, and $870$.
Other methods were also tested for comparison, including the direct application of a gradient-free optimizer, the direct utilization of gradient descent, and the Snake algorithm. We note that the trained evaluator is also necessary for the traditional methods, because conducting real experiments would require significant time and resources~\cite{Introsnake_2}, and numerical simulation for large-scale SQECs is also challenging. 
All four methods employ the same trained evaluator on the same i9-14900KF CPU. Fig.~\ref{fig2}(a) demonstrates that the well-trained GNNs-based designer achieves lower crosstalk errors than all other methods for all the graph scales. Additionally, the GNNs-based designer is exponentially faster than all other methods for all given tasks, as shown in Fig.~\ref{fig2}(b). Especially for large-scale graphs where direct optimizations are not feasible, the GNNs-based designer achieves slightly more than half the errors of the Snake algorithm in several seconds, while the Snake algorithm would take dozens of minutes.


\textit{Discussion--} This study demonstrates the advantages of applying GNNs to the hardware-level optimization of quantum chips for the first time, especially for large-scale ones.
We apply two graph-based neural-network models named evaluator and designer to tackle the challenge of mitigating quantum crosstalk errors in large-scale SQECs.
The evaluator can estimate crosstalk for large-scale SQECs containing thousands of qubits.
And the designer outperforms traditional optimization algorithms in efficiency, effectiveness, and scalability.
In the future research, additional error sources and variations between qubits may be taken into account, and the evaluator may be trained using experimental data, thus the practicality of the proposed algorithm can be further enhanced.


\textit{Acknowledgments--}We acknowledge J.-L. Long, Z. Wang, R.-B. Wu, G.-Z. Zhu, Y. Shang, T.-T. Chen, and G.-S. Liu for discussions. This work was support by Innovation Program for Quantum Science and Technology (Grant No. 2021ZD0300200), and the National Natural Science Foundation of China with Grant No. 92365209.

\textit{Code Availability--}The supporting code, data, and trained models for this study are openly available from GitHub~\cite{github}.


\bibliography{ref}

\end{document}


\title{Supplementary Materials of \\
Scalable Parameter Design for Superconducting Quantum Circuits with Graph Neural Networks}

\author{Hao Ai}
\affiliation{School of Integrated Circuits, Tsinghua University, Beijing 100084, China.}
\author{Yu-xi \surname{Liu}}
\altaffiliation[yuxiliu@mail.tsinghua.edu.cn]{}
\affiliation{School of Integrated Circuits, Tsinghua University, Beijing 100084, China.}
\date{\today}

\maketitle
\renewcommand{\thefigure}{S\arabic{figure}}
\renewcommand{\thetable}{S\arabic{table}}

\section{Algorithm Details}

\subsection{Evaluator: Graph-Based MLPs}\label{evaluator_appendix}

\begin{figure*}[htbp]
	\centering
	\includegraphics[width=1\linewidth]{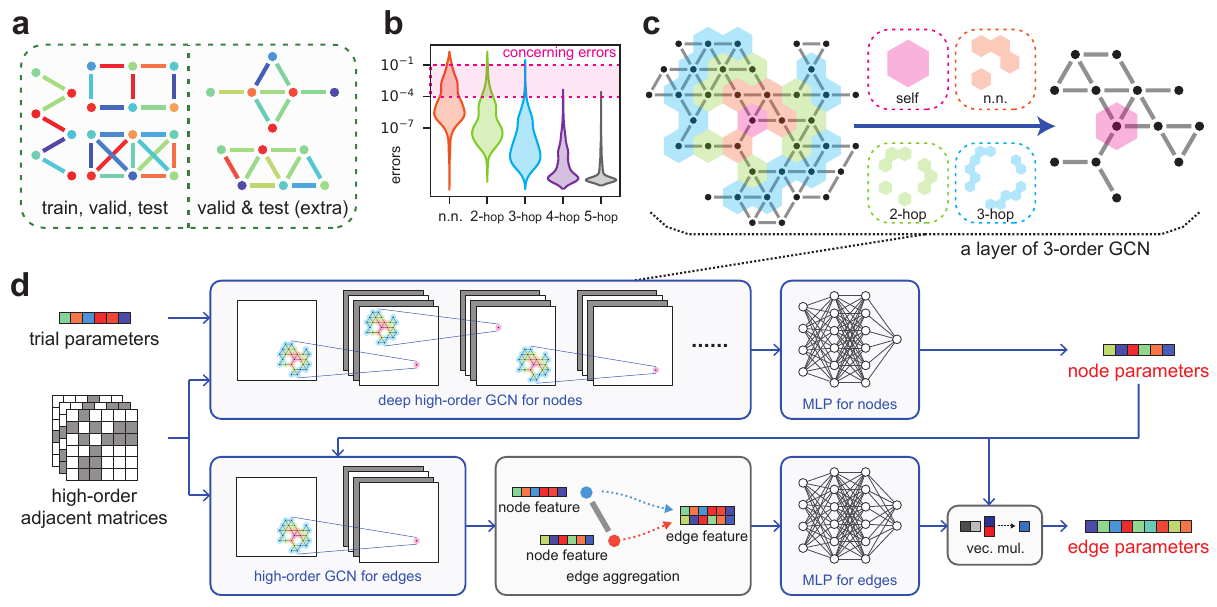}\\
	\caption{
		\textbf{Details for the proposed algorithm.}
		\textbf{a.} The small-scale graphs for training, validating, and testing the evaluator, which are the most common qubit-coupling structures. The node and edge parameters are randomly sampled to generated the datasets. 
		Each graph contains $6$ nodes, the left $3$ graphs are used for training, validating, and testing.
		While the right $2$ graphs are used for extra validating and testing, guaranteeing the evaluator's generalization ability to different underlying structures.
		\textbf{b.} The exponentially decay of the interplay between two nodes with their distance.
		Dataset for plotting \textbf{b} is obtained by simulating single-qubit $R_X$ operations on a small-scale graph.
		The horizontal axis represents the distance between nodes, and the vertical axis represents the error in logarithmic coordinates.
		The width of the plot at different values indicates the amount of the data.
		Since only the errors ranging from $10^{-4}$ to $10^{-1}$ are concerned, the crosstalk errors between nodes with far distances can be neglected. This guarantees the evaluator's generalization ability to larger-scale graphs.
		\textbf{c.} A layer of 3-order GCN. The node features of itself (magenta), its nearest neighbor nodes (orange), its $2$-hop nodes (green), and its $3$-hop nodes (blue) are aggregated.
		\textbf{d.} The detailed architecture of the designer. High-order adjacent matrixes and trial node parameters are input to the designer. First, a deep high-order GCN and an MLP can assign parameters for each node, which constitute part of the designer's output. Second, these node parameters serve as inputs for another high-order GCN to output new node features, which are then aggregated by edges and fed into another MLP. The MLP outputs each edge parameter's selection preferences $p_i,p_j$ from its two nodes' parameters $\omega_i,\omega_j$. Then vector multiplication (vec. mul.) is performed to obtain the corresponding edge parameter, $\omega_{ij} = p_i \omega_i + p_j \omega_j$. Thus the edge parameters are assigned to constitute another part of the designer's output. In a word, the designer can design both node and edge parameters simultaneously.
	}\label{figs1}
\end{figure*}

As an essential property making the algorithm effective, the generalization ability of the evaluator is guaranteed in two ways. First, extra graphs only for testing purposes are introduced as depicted in Fig.~\ref{figs1}(a). Each graph contains $6$ nodes, the left $3$ graphs are used for training, validating, and testing, while the right $2$ graphs are used for extra validating and testing. The extra testing graphs ensure that the trained evaluator can be applied to graphs with new underlying structures (NS). Note that the results of validating on new structures are not shown in the main text for simplicity.
Second, the influence of the parameters in question should be localized. In other words, for two nodes, e.g., $\mathcal{N}_i$ and $\mathcal{N}_k$ in a graph, the undesired impact from $\mathcal{N}_i$ to $\mathcal{N}_k$ decays exponentially with the distance between the two nodes, as illustrated in Fig.~\ref{figs1}(b). Dataset for plotting Fig.~\ref{figs1}(b) is obtained by simulating single-qubit $R_X$ operations on a small-scale graph. The localized nature of the parameters is essential for the evaluator, which ensures that the MLPs trained on small-scale graphs remain effective for larger-scale ones. The trained evaluator is also tested on a linear-coupled graph with $30$ qubits (30Q), ensuring the evaluator's generalization ability to larger graphs.

The evaluator comprises multiple MLPs to estimate the errors associated with different types of quantum operations. Each MLP is tailored to a specific type of quantum operations, such as $R_X$, $R_Y$, or $R_{XY}$ operations.
These MLPs are trained to output the base-$10$ logarithm of respective crosstalk errors, while the inputs for single- and two-qubit operations are slightly different.
The MLPs for single-qubit operations are trained to predict the undesired excitations from one node to another. Taking the average crosstalk error of $R_X$ operation from node $\mathcal{N}_i$ to node $\mathcal{N}_k$ as an example, the MLP takes the node frequencies $\omega_i,\omega_k$ and the corresponding matrix elements $(A^p)_{ik}$ as inputs, where $(A^p)_{ik}$ denotes the element of the $p$-th order adjacency matrix as defined before.
The MLP for two-qubit $R_{XY}$ operations can evaluate the crosstalk error from edge $\mathcal{E}_{ij}$ to node $\mathcal{N}_k$, where $k\neq i,j$. Similar to the case of single-qubit operations, the inputs for two-qubit operations encompass the node frequencies $\omega_i,\omega_j,\omega_k$, the edge frequency $\omega_{ij}$, as well as the matrix elements $(A^p)_{ik}$ and $(A^p)_{jk}$.
Each MLP comprises $4$ hidden layers, each with $32$ neurons. The detailed architecture of the evaluator is shown in Figs.~\ref{figs9}(a) and (b).

\begin{figure*}[htbp]
	\centering
	\includegraphics[width=1\linewidth]{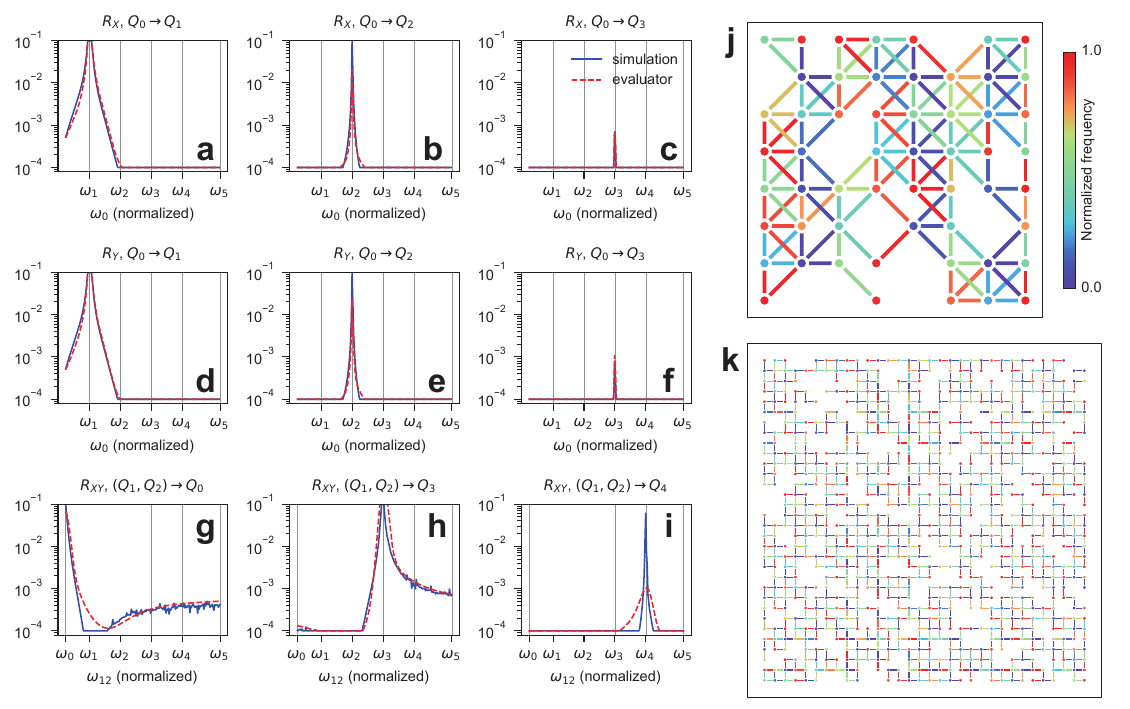}\\
	\caption{
		\textbf{Demonstrations of the trained evaluator and designer.}
		\textbf{a-i.} Some examples of fitted and simulated crosstalk errors between neighboring qubits in a linear graph with $6$ qubits numbered sequentially '$Q_0-Q_1-Q_2-Q_3-Q_4-Q_5$'. The horizontal axes represent frequencies of nodes or edges where quantum operations are applied, and the vertical axes represent the corresponding crosstalk errors in logarithmic coordinates. The results of numerical simulation are plotted with the solid blue lines, while the outputs of the trained evaluator are depicted in the dashed red lines.
		The peaks of the blue or red lines represent the maximum errors.
		The gray vertical lines indicate the idling frequencies of the qubits. The fitted and simulated results agree well with each other, showcasing the effectiveness of the evaluator.
		\textbf{j-k.} Examples of the frequency assignments designed by the trained designer for a medium-scale graph and a large-scale graph. The node and edge colors correspond to the normalized node and edge frequencies. These figures demonstrate the designer's effectiveness for considering nearest and higher-hop neighbors for both nodes and edges.
	}\label{figs2}
\end{figure*}

To demonstrate the reliability of the trained evaluator, we select various frequency assignments on a linear graph with $6$ qubits numbered sequentially '$Q_0-Q_1-Q_2-Q_3-Q_4-Q_5$', and compare the numerical simulation results with the fitting results from the evaluator. In Figs.~\ref{figs2}(a)-(i), the solid blue lines describe the results from numerical simulations, while the dashed red lines represent the fitting results from the evaluator, and the two results agree well with each other.
For single-qubit operations, we vary the node frequency $\omega_0$ of qubit $Q_0$ while maintaining the other frequencies constant. The crosstalk excitations on qubits $Q_1,\,Q_2$ and $Q_3$ are plotted to signify the crosstalk errors induced by single-qubit operations.
For two-qubit operations, we scan the edge frequency of coupled-qubit pair $(Q_2,Q_3)$ with the other frequencies unchanged, and plot the crosstalk excitations on qubits $Q_1,\,Q_4$ and $Q_5$.
The fitted and simulated results agree well with each other, showcasing the effectiveness of the evaluator.

\begin{figure}[tbp]
	\centering
	\includegraphics[width=1\linewidth]{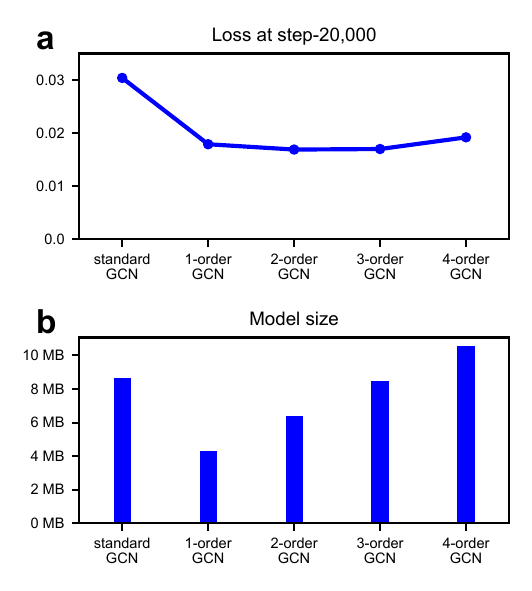}\\
	\caption{
		\textbf{Comparison between standard and high-order GCNs.}
		\textbf{a.} The training losses at the $20,000$-th step.
		\textbf{b.} The model sizes of compared GCNs. High-order GCNs perform much better than standard GCN with similar model sizes. This comparison demonstrates the necessity of introducing high-order GCNs, rather than simply using standard GCN.
		In this study, three-order GCN is used because the dominant errors only relate to the nearest neighboring, $2$-hop neighboring and $3$-hop neighboring qubits, as shown in Fig.~\ref{figs1}(b).
	}\label{figs3}
\end{figure}

\begin{figure}[htbp]
	\centering
	\includegraphics[width=1\linewidth]{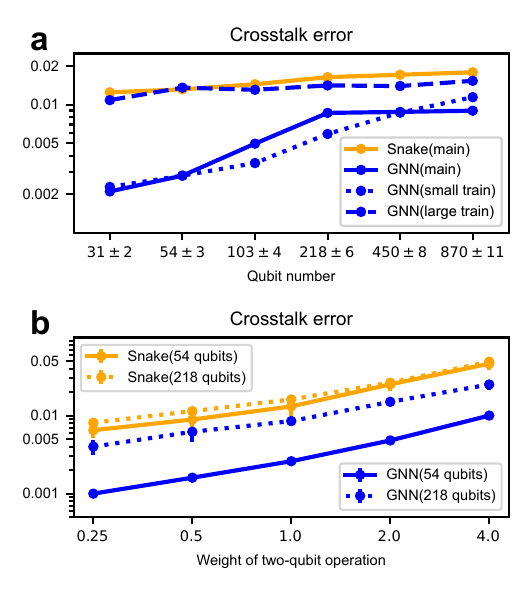}\\
	\caption{
		\textbf{a.} Results of different training scales for the designer. The solid blue line represents the results in the main text, corresponding to the designer trained with the default $54$, $103$, and $218$ qubits. The two blue dashed lines represent the designer trained with smaller and larger graphs, respectively. Memory usage is controlled similarly. The solid orange line indicates the results of the Snake algorithm, also from the main text, serving as a reference.
		\textbf{b.} Results of different weights assigned to two-qubit operations. The horizontal axis denotes the error weights for two-qubit operations, while the vertical axis shows errors on a logarithmic scale. The blue lines indicate results from the GNNs-based designer, and the orange lines correspond to the Snake algorithm. Solid lines indicate tests with $54$ qubits, and dashed lines indicate tests with $218$ qubits. In all scenarios, the GNNs-based designer consistently achieves lower errors.
	}\label{figs4}
\end{figure}

\subsection{Designer: GNNs and MLPs}\label{designer_appendix}

For the designer, considering the errors' dependence on higher-order neighboring parameters, we apply higher-order graph convolutional networks (GCNs) to solve the problem. The architecture of high-order GCNs used here is inspired by some similar GNNs~\cite{ALG_HighOrderGCN,ALG_HighOrderGCN2}. As exemplified in Fig.~\ref{figs1}(c), a layer of higher-order GCN is defined as
\begin{equation}
	F_{(l+1)} = \sigma\left( F_{(l)} W^0_{(l)} + \tilde{A}^1 F_{(l)} W^1_{(l)} + \tilde{A}^2 F_{(l)} W^2_{(l)} + \cdots \right),
\end{equation}
where $\tilde{A}^p = (\tilde{D}^p)^{-1/2} (A^p + I) (\tilde{D}^p)^{-1/2}$ (with $p=0,1\,,2,\cdots$) stands for the normalized $p$-th order adjacency matrix, where $I$ is the identity matrix and $(\tilde{D}^p)_{ik} = \sum_k (A^p + I)_{ik}$. $(A^p)_{ik}=1$ if the distance of $\mathcal{N}_i$ and $\mathcal{N}_k$ is $p$, otherwise, $(A^p)_{ik}=0$. $F_{(l)}$ is the feature matrix of the $l$-th layer, $W^p_{(l)}$ is the trainable weight, and $\sigma(\cdot)$ is the nonlinear activation function. 

As shown in Fig.~\ref{figs1}(d), the designer is used to design parameters on both nodes and edges within the input graphs.
Given a batch of graphs and random node frequencies as inputs, a deep high-order GCN and an MLP can assign frequencies for each node, which constitute part of the designer's output. Subsequently, these node frequencies serve as inputs for another high-order GCN to output new node features, which are then aggregated by edges and fed into an MLP. For an edge $\mathcal{E}_{ij}$, with a softmax function applied to the two output-layer neurons, the MLP generates the selection preferences $p_i$ and $p_j$ for the respective edge frequency $\omega_{ij}$ from its two node frequencies $\omega_i$ and $ \omega_j$. Ultimately, each edge's frequency is determined by
\begin{equation}
	\omega_{ij} = p_i \omega_i + p_j \omega_j.
\end{equation}
Thus another part of the designer's output is obtained.
In this work, the three-order GCN for node frequencies comprises $12$ hidden layers, and contains $256$ neurons for most of the layers. While the three-order GCN for edge frequencies consists of $3$ hidden layers, each with $64$ neurons. As a comparison, the training effects and model sizes of standard GCN and different high-order GCNs are plotted in Fig.~\ref{figs3}. High-order GCNs perform much better than the standard one, demonstrating the necessity of introducing high-order GCNs. Three-order GCN is used in the designer because the relevant errors mainly relate to the nearest neighboring, $2$-hop neighboring and $3$-hop neighboring qubits, as shown in Fig.~\ref{figs1}(b). The detailed architecture of the designer is shown in Figs.~\ref{figs9}(c) and (d).
The complete form of designer's loss is
\begin{equation}\begin{aligned}
		\mathcal{L}^{\rm designer} (\mathcal{G},\vec\omega) = & \frac{w_X}{|\mathcal{N}|} \sum_{\mathcal{N}_i\in\mathcal{N}} \sum_{k\neq i} \hat{E}_{X,i\to k} \\
		+ \frac{w_Y}{|\mathcal{N}|} \sum_{\mathcal{N}_i\in\mathcal{N}}& \sum_{k\neq i} \hat{E}_{Y,i\to k} + \frac{w_{XY}}{|\mathcal{E}|} \sum_{\mathcal{E}_{ij}\in\mathcal{E}} \sum_{k\neq i,j} \hat{E}_{{XY},{ij}\to k}, \label{eq:3}
\end{aligned}\end{equation}
First, the estimated crosstalk errors, e.g., $\hat{E}_{X,i\to k}$, on different nodes $\mathcal{N}_k$ are summed to obtain the total crosstalk error of operations on node $\mathcal{N}_i$ (or edge $\mathcal{E}_{ij}$). Second, the node(edge)-related errors are subsequently averaged to get the errors for one specific type of quantum operations. Then, the crosstalk errors of all three types of quantum operations are summed up to obtain the comprehensive quantum crosstalk error, serving as the loss function $\mathcal{L}^{\rm designer} (\mathcal{G},\vec\omega)$ under the specific graph $\mathcal{G}$ and frequency assignment $\vec\omega$. Finally, a batch of crosstalk errors for different graphs, together with the parameter assignment on each graph, i.e., $(\mathcal{G}, \vec\omega)$, are averaged to calculate the total loss for unsupervisedly training the designer.
In our simulation, only the crosstalk errors up to 4-hop neighbors are taken into account, as the crosstalk errors diminish with distance.

When applying the designing on a specific graph, the batched utilization strategy can further enhance the designer's effectiveness, as mentioned in the main text. The graphs with average qubit numbers of $32$, $54$, $103$, $218$, $450$ are tested with batch sizes of $512$, $512$, $512$, $64$, $8$, and $2$, respectively.
Figs.~\ref{figs2}(j) and (k) show the parameter assignments output by the trained designer for medium- and large-scale graphs. The node and edge colors correspond to the normalized node and edge frequencies. These figures demonstrate the effectiveness of considering nearest and multi-hop neighbors for both nodes and edges, ensuring that their frequencies are as distinct as possible.

We explore the impact of the training graph scale on the designer's performance, as illustrated in Fig.~\ref{figs4}(a). All test settings are consistent with those in Fig.~2 of the main text. The solid blue line indicates the results of training the designer on graphs with $54$, $103$, and $218$ nodes, which serves as the default setting in the main text. The two blue dashed lines correspond to training on smaller ($31$, $54$, $103$ nodes) and larger ($103$, $218$, $450$ nodes) graphs, respectively. In all the three training scenarios, the memory usage is controlled approximately constant, while the corresponding batch sizes and the total number of nodes per training step are shown in Tab.~\ref{tables_trainingscales}. Training on larger-scale graphs performs worse due to the smaller batch size and total node number in each step. As for the other two scenarios, the designer trained on smaller graphs performs better in tests with $103$ and $218$ nodes, while the default setting performs better in the $870$-qubit test. For comparison, the results of the Snake algorithm are shown as the orange line.

\begin{table}[tbp]
	\caption{\label{tables_trainingscales} Different training scales of the designer}
	\begin{ruledtabular}\begin{tabular}{lrrr}
			
			& smaller & default & larger \\
			\colrule
			batch size 
			& 1,530 & 288 & 45 \\
			nodes per graph
			& 31,54,103 & 54,103,218 & 103,218,450 \\
			total nodes per step
			& 31,960 & 12,000 & 3,855 \\
			memory usage\footnote{Controlled variable}
			& $\sim 19$GB & $\sim 19$GB & $\sim 19$GB 
\end{tabular}\end{ruledtabular}\end{table}

We also test the robustness of the designer against different two-qubit operation error weights. In the main text, the error weights for single-qubit and two-qubit operations are both set to $1$. In Fig.~\ref{figs4}(b), we present the results of the GNNs-based designer and the Snake algorithm under varying two-qubit operation error weights, while keeping the single-qubit operation error weight fixed at $1$. Note that these tests do not involve retraining the designer; instead, we use the designer trained with the $1:1:1$ weights in the main text. We conduct tests on graphs with an average of $54$ and $218$ qubits. The results of the Snake algorithm are shown as orange lines, while the designer's results are depicted as blue lines. It can be observed that the GNNs-based designer achieves lower errors than the Snake algorithm across all weights and qubit numbers.

\subsection{Traditional Methods}\label{tradional_methods}

\begin{figure*}[tbp]
	\centering
	\includegraphics[width=1\linewidth]{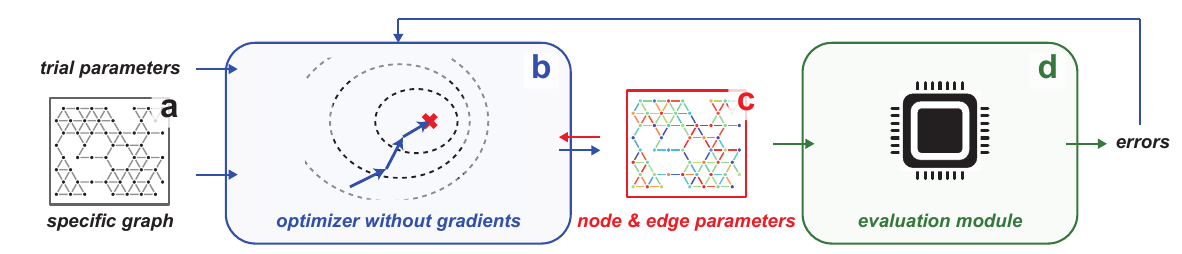}\\
	\caption{
		\textbf{Direct optimization algorithms.} For a specific graph, randomly generated trial parameters for all the nodes and edges are input into the optimizer. The optimizer iteratively explores new parameter assignments, gives them to the evaluation module for evaluating the corresponding errors, and then feeds these errors back to ultimately obtain an appropriate parameter assignment.
		\textbf{a.} An example of the specific graph.
		\textbf{b.} The gradient-free optimizer.
		\textbf{c.} Node and edge parameters designed for the input graph.
		\textbf{d.} The evaluation module, e.g., real experiments for small-scale SQECs, or the trained evaluator for larger SQECs.
	}\label{figs5}
\end{figure*}

In addition to the GNNs-based designing algorithm proposed in this study, three traditional methods are also discussed for comparisons. These include the direct utilization of a gradient-free optimization algorithm (DirectOptim), the direct application of gradient descent algorithm (GradAlg), and the Snake algorithm for large-scale graphs~\cite{Introsnake_1, Introsnake_2}.
\begin{enumerate}
	\item[1.] DirectOptim: A straightforward consideration for the parameter designing problem is to utilize feedback-based optimization algorithms~\cite{scipy} to find available parameter assignments for specific coupled-qubit graphs directly, as shown in Fig.~\ref{figs5}.
	For a specific graph as exemplified in Fig.~\ref{figs5}(a), randomly generated trial parameters for all the nodes and edges are input into the optimizer. The optimizer, shown in Fig.~\ref{figs5}(b), iteratively explores new parameter assignments, such as Fig.~\ref{figs5}(c). In Fig.~\ref{figs5}(d), the assignments are given to the evaluation module to assess the corresponding errors. The evaluation module can utilize real experiments for small-scale SQECs or employ the trained evaluator for larger-scale SQECs. Then the errors are fed back to the optimizer to ultimately obtain an appropriate parameter assignment.
	\item[2.] GradAlg: The direct application of gradient descent algorithm is similar to DirectOptim, which replaces the gradient-free optimizers in DirectOptim with gradient descent algorithms. Since the errors under given frequency assignments can be estimated through the evaluator, the corresponding gradient information is available after the errors are calculated. So, gradient descent algorithms can be utilized to find the frequency assignments with mitigated errors.
	\item[3.] Snake: Due to the low computational efficiency, DirectOptim and GradAlg are not applicable for large-scale SQECs. Snake algorithm was proposed as an improvement of direct optimization by dividing the entire graph into multiple small-scale subgraphs and sequentially applying the optimizer to optimize these subgraphs~\cite{Introsnake_1,Introsnake_2}. Thus, Snake algorithm can be used to design parameters for larger-scale graphs, such as the Sycamore processor, which is said to be the first to achieve quantum supremacy~\cite{IntroQuRealize_SQC_google2019}.
\end{enumerate}
All these traditional algorithms overlook the graph structures established by the qubits and their couplings. This oversight may significantly impact the effectiveness and efficiency of optimization when considering large-scale graphs.

In our tests, both GNNs-based designing algorithm and the three traditional algorithms leverage the trained evaluator to estimate the crosstalk errors.
For Snake and DirectOptim, Powell optimizer is employed~\cite{Powell}, and all other settings adhere to the default settings provided by Scipy~\cite{scipy}. The scope of Snake is $4$, which is the best choice in Ref.~\cite{Introsnake_2}.
Among these four algorithms, the GNNs-based designer and Snake are available for large-scale graphs, whereas the other two are limited to medium-scale graphs due to their inefficiency.
On the other hand, the GNNs-based designer and GradAlg require gradient information from the evaluator, while the other methods can use gradient-free optimizers.
Following a prolonged training phase, the GNNs-based designer can be swiftly applied to graphs of varying structures and scales. In contrast, the traditional methods necessitate distinct optimization for each specific graph with relatively slower processing speeds.

\section{Demonstration with Quantum Crosstalk}

\begin{figure*}[tbp]
	\centering
	\includegraphics[width=1\linewidth]{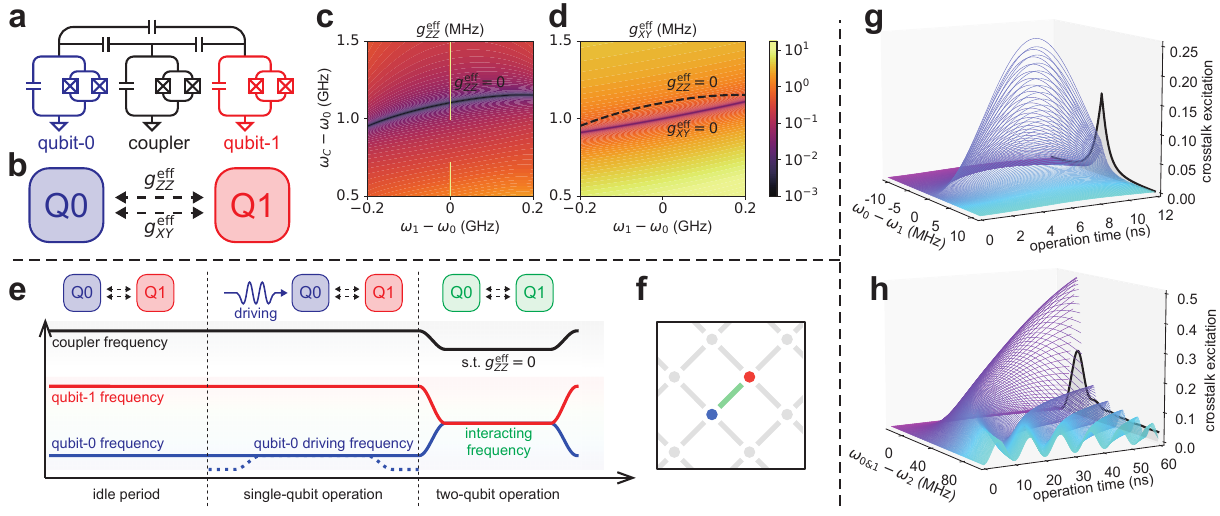}\\
	\caption{
		\textbf{The physical system, quantum operations, and crosstalk analysis.}
		\textbf{a.} The interconnected 'qubit-coupler-qubit' subsystem as the building block of SQECs.
		\textbf{b.} Effective $ZZ$ and $XY$ couplings between the qubits, resulting in different types of quantum crosstalk. The coupler mode is equivalently eliminated.
		\textbf{c.} The variations of effective $ZZ$ coupling strength $g_{ZZ}^{\rm eff}$ as a function of the qubit and coupler frequencies. The dark line where $g_{ZZ}^{\rm eff}=0$ is referred to as the $ZZ$ free line.
		\textbf{d.} The variations of the effective $XY$ coupling strength $g_{XY}^{\rm eff}$ as a function of the qubit and coupler frequencies. The fitted $ZZ$-free line is depicted with the black dashed line.
		\textbf{e.} The control schemes and the corresponding frequency configurations for idle periods and different operations, where different frequencies are represented by different colors. In the idle period, the frequencies corresponding to the qubits and the coupler are $\omega_{0}$, $\omega_{1}$ and $\omega_{C}$, which are chosen to ensure that the effective couplings are negligibly small or minimized. For single-qubit operations on qubit $Q_0$, $Q_0$ is resonantly driven by external fields for a certain period of time (dashed blue line). For two-qubit operations, the qubit frequencies are adjusted to to enable resonant interaction for a specific duration. Meanwhile, the coupler frequency is adjusted to make $g_{ZZ}^{\rm eff}=0$, which may result in a non-zero value of $g_{XY}^{\rm eff}$.
		\textbf{f.} The graph-coloring representation of frequency assignments for a building block (blue and red dots connected by a green segment) of large-scale SQECs (lattice with the gray color). The node frequencies correspond to the qubit frequencies during the idle periods and single-qubit operations, and the edge frequencies denote the resonant frequencies that two qubits resonantly interact to realize two-qubit operations.
		\textbf{g.} The crosstalk excitations on qubit $Q_1$ when single-qubit operations are implemented on qubit $Q_0$ in building block of SQECs.
		\textbf{h.} The crosstalk excitations on qubit $Q_2$ when two-qubit operations are implemented on $Q_0$ and $Q_1$ in a system of three qubits with two couplers. Because operation times are related to the parameters given by the users, the errors are averaged over multiple operating times and depicted with the black lines. The peaks correspond to the maximum errors.}\label{figs6}
\end{figure*}

\subsection{Physical System}\label{physical_system}

The building block of SQECs is the interconnected 'qubit-coupler-qubit' subsystem~\cite{coupler} as schematically shown in Fig.~\ref{figs6}(a), where three frequency-adjustable transmons~\cite{transmon} are capacitively coupled in pairs. The middle one functions as the tunable coupler, which is now extensively utilized in SQEC experiments, e.g. in Refs.~\cite{IntroQuRealize_SQC_google2019,IntroQuRealize_SQC_zhuxiaobo2021}.
The building-block Hamiltonian of SQEC is
\begin{equation}\begin{aligned}\label{eq_hamiltonian}
		H = &\sum_{i\in Q\cup C} \hbar\omega_i(t) a_i^\dagger a_i + \sum_{i\in Q\cup C} \hbar\frac{\alpha}{2} a_i^\dagger a_i \left(a_i^\dagger a_i -1\right) \\
		&+ \sum_{i,j\in Q\cup C} \hbar g_{ij} \left( a_i^\dagger a_j + a_i a_j^\dagger \right) \\
		&+ \sum_{i\in Q} \hbar \Omega_i^x(t) \cos[\omega_{di}^x(t)t] \left( a_i^\dagger + a_i \right) \\
		&-i \sum_{i\in Q} \hbar \Omega_i^y(t) \sin[\omega_{di}^y(t)t] \left( a_i^\dagger - a_i \right) ,
\end{aligned}\end{equation}
where $\hbar$ is the reduced Plank's constant. The set $C$ only includes the middle transmon acting as the coupler, while the set $Q$ includes other two transmons as the qubits. $\omega_i(t)$ is the eigen-frequency of the first excitation, $a_{i}$ and $a^\dagger_{i}$ are the annihilation and creation operators of the $i$-th transmon, respectively. We assume that all three transmons have the same anharmonicity $\alpha$ for simplicity of discussions. The parameter $g_{ij}$ denotes the coupling strength between the $i$-th and the $j$-th transmons, $\omega_{di}^{x}(t)$ and $\omega_{di}^{y}(t)$ are the driving frequencies corresponding to the driving strengths $\Omega_i^{x}(t)$ and $\Omega_i^{y}(t)$ for different single-qubit operations~\cite{IntroQuRealize_Gu2017}. We note that the time-dependent terms can be controlled or devised by the experimenters.

In this building block, the two qubits typically have the resonant or near resonant frequencies, while the middle coupler is often detuned significantly from two qubits. Thus, the coupler mode can be equivalently eliminated, leading to the weak effective $ZZ$ and $XY$ couplings between two qubits~\cite{Crosstalk_zhaopeng}, as illustrated in Fig.~\ref{figs6}(b).
According to the definition~\cite{Crosstalk_zhaopeng,zz_coupling}, the $ZZ$ coupling can be given as
\begin{equation}
	g_{ZZ}^{\rm eff} \equiv (E_{\ket{110}} - E_{\ket{100}}) - (E_{\ket{010}} - E_{\ket{000}}),
\end{equation}
where $E_{\ket{Q0,Q1,C}}$ represents the energy corresponding to the state  $\ket{Q0,Q1,C}$, in which qubit $Q_0$, qubit $Q_1$ and the coupler are in the states $|Q0\rangle$, $|Q1\rangle$ and $|C\rangle$, respectively. The variations of the effective $ZZ$ coupling strength $g_{ZZ}^{\rm eff}$ via the frequencies of the qubits and coupler can be numerically solved by diagonalizing the Hamiltonian without driving fields, as depicted in Fig.~\ref{figs6}(c).
By taking the Schrieffer-Wolff transformation~\cite{SW_transfomation}, the effect of the coupler can also be equivalent to the effective $XY$ coupling,
\begin{equation}
	g_{XY}^{\rm eff} = g_{QQ} - \frac{g_{QC}^2}{2} \left( \frac{1}{\omega_C - \omega_0} + \frac{1}{\omega_C - \omega_1} \right),
\end{equation}
where $\omega_0,\omega_1,\omega_C$ are the frequencies of $Q_0$, $Q_1$ and coupler, $g_{QQ}$ is the direct coupling strength between two qubits, and $g_{QC}$ is the direct coupling strength between the qubits and the coupler. $g_{XY}^{\rm eff}$ is plotted in Fig.~\ref{figs6}(d).
The existence of non-zero $ZZ$ and $XY$ couplings results in different types of quantum crosstalk errors, however, as shown in Figs.~\ref{figs6}(c) and (d), the non-simultaneous-zero values of the two couplings present a theoretical challenge in completely eliminating crosstalk errors between the two qubits.
To address this challenge, we adjust $\omega_C$ to keep the system on the so-called $ZZ$-free line (i.e., $g_{ZZ}^{\rm eff}=0$)~\cite{Crosstalk_zhaopeng}, aiming to eliminate the crosstalk errors induced by the $ZZ$ coupling. Consequently, qubit frequencies can be designed to mitigate crosstalk errors arising from the $XY$ coupling.

\subsection{Quantum Operations}\label{quantum_operation}

Quantum crosstalk commonly arises during quantum operations, also called as quantum gates. We select single-qubit $R_X$, $R_Y$~\cite{NielsenBook}, and two-qubit $R_{XY}$ operations~\cite{iSWAP} as the fundamental quantum gates. According to the quantum computation principles, these three types of operations can be combined to construct universal quantum computation~\cite{NielsenBook}.
Fig.~\ref{figs6}(e) illustrates the control for the frequencies of qubits, coupler and driving fields across idle periods, single-qubit operations, and two-qubit operations. Here, the idle periods denote that no quantum operations are applied~\cite{Crosstalk_zhaopeng}.
\begin{enumerate}
	\item[1.] During idle periods, $\omega_0$ and $\omega_1$ are carefully assigned to minimize crosstalk errors arising from the effective $XY$ coupling, while $\omega_C$ is positioned on the $ZZ$-free line to ensure that crosstalk errors induced by the effective $ZZ$ coupling are eliminated.
	\item[2.] When operating a single-qubit operation on qubit $Q_0$, the frequencies $\omega_0$, $\omega_1$ and $\omega_C$ remain unchanged, while a resonant driving field with $\omega_{0d}=\omega_0$ is applied to $Q_0$, as denoted by the blue dashed line in Fig.~\ref{figs6}(e). By regulating the intensity, duration, and polarization direction (e.g., $X$ or $Y$) of the driving field, a specific single-qubit operation on $Q_0$ can be achieved.
	\item[3.] During a two-qubit operation involving $Q_0$ and $Q_1$, both qubits are adjusted to have the same frequency, resonantly interact for a specific duration to realize a specific operation. Simultaneously, $\omega_C$ is adjusted in synchronization to ensure that the system consistently operates on the $ZZ$-free line.
\end{enumerate}
Fig.~\ref{figs6}(f) shows the corresponding graph-coloring~\cite{GCP_1st,GCP_amazon} representation of the frequency assignment, where the frequencies are normalized and represented by different colors. The node frequencies correspond to the qubit frequencies during the idle periods and single-qubit operations, and the edge frequencies denote the resonant frequencies that two qubits resonantly interact to realize two-qubit operations. Both node and edge frequencies constitute the parameters to be designed by the algorithm proposed in this study.

\subsection{Crosstalk Analysis}\label{crosstalk_analysis}

In order to quantitatively describe quantum crosstalk, the undesired excitations occurring on a specific qubit $Q_k$ are simulated when single-qubit (or two-qubit) operations are applied to qubit $Q_i$  (or coupled-qubit pair $Q_i\&Q_j$) with $k\neq i,j$.
We define such excitations on $Q_k$ as crosstalk errors from node $\mathcal{N}_i$ (or edge $\mathcal{E}_{ij}$) to node $\mathcal{N}_k$. In our simulations, $Q_i$ is (or $Q_i\&Q_j$ are) initialized to some excited states, while $Q_k$ is set in its ground state. After finishing the simulation, the average excitation on $Q_k$ is calculated to reflect the crosstalk error under the corresponding configurations.

For a given type of quantum operations, the main factors influencing crosstalk errors include the qubit frequencies, the qubit distances within the graph, the inter-qubit coupling strengths, the driving strengths of single-qubit operations, and the time durations of operations.
\begin{enumerate}
	\item[1.] Qubit frequencies: Qualitatively, because the conservation of both energy and excitation number must be satisfied simultaneously, the quantum crosstalk error between two qubits is expected to increase as their frequency difference decreases. As shown in the black lines in Figs.~\ref{figs6}(g) and (h), the correlations between nearest-neighbor crosstalk errors and frequency differences are evident, which facilitates the fitting with neural networks.
	\item[2.] Qubit distances within the graph: Given that the distant qubits can only interact through weak indirect or parasitic couplings~\cite{nnn_coupling}, the quantum crosstalk error between them would significantly decrease with the increase of their distance, as exemplified in Fig.~\ref{figs1}(b). Qubit distances are also taken into account by the evaluator to estimate quantum crosstalk errors.
	\item[3.] Coupling and driving strengths: The coupling strengths between two qubits (or between a qubit and a coupler) and the driving strengths for single-qubit operations can also influence quantum crosstalk errors.
	The decrease of the coupling and driving strengths generally helps to mitigate crosstalk errors, but this often impacts the efficiency of quantum operations~\cite{Crosstalk_zhaopeng}. Moreover, the trade-off and adjustment of coupling and driving strengths are independent on the graph structure, thus their values are fixed in this study.
	\item[4.] Time durations of operations: As demonstrated in Figs.~\ref{figs6}(g) and (h), time durations could affect crosstalk excitations when single- or two-qubit operations are implemented. Nevertheless, when other parameters are given, the time duration for a specific operation is determined by the user-input parameters, such as the rotation angle of $R_X$ gate. Therefore, crosstalk errors are averaged over multiple time durations in our numerical simulations.
\end{enumerate}

In this work, classical resources of errors are not considered.
For example, defects in fabrication~\cite{defects} and distortions in modulating microwave pulses~\cite{pulse} can also affect the quality of quantum information processing.
Additionally, when signals are applied through specific control lines, the microwave current induced magnetic fields may introduce unwanted signals in other control lines, leading to what is known as classical crosstalk. Many works are studying the mitigation of classical crosstalk errors~\cite{Crosstalk-classical,Crosstalk_classical2,Crosstalk_classical_GGP}. The classical resources of errors are not taken into account in this study, because they are usually graph-independent.

In a word, after the coupling and driving strengths are given and  multiple operating time durations are averaged, the quantum crosstalk errors only depend on the node and edge frequencies, graph structures, and operation types. Thus, neural networks can be trained to evaluate the frequency assignments for quantum crosstalk mitigation.

\begin{figure*}[htbp]
	\centering
	\includegraphics[width=1\linewidth]{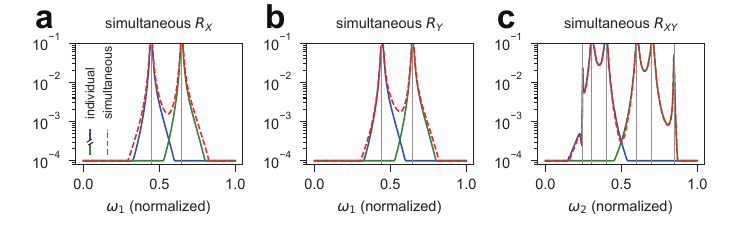}\\
	\caption{
		\textbf{a-c. Demonstration of simultaneous operations}. The crosstalk excitations caused by simultaneous operations, as shown with the dashed red lines, are approximately the sum of those induced when applying the corresponding operations individually, plotted with the solid lines.
	}\label{figs7}
\end{figure*}

\subsection{Numerical Simulations}\label{numerical_simulation}

To simulate quantum operations, the Hamiltonian in Eq.~(\ref{eq_hamiltonian}) is transformed into a rotating frame with the rotating wave approximations~\cite{RWA}. Taking the single-qubit $R_X$ operation applying on qubit $Q_0$ as an example, the transformed Hamiltonian without other operations in Eq.~(\ref{eq_hamiltonian}) is given as
\begin{equation}\begin{aligned}
		\tilde{H} = &\sum_{i\in Q\cup C} \hbar(\omega_i - \omega_0) a_i^\dagger a_i + \sum_{i\in Q\cup C} \hbar\frac{\alpha}{2} a_i^\dagger a_i \left(a_i^\dagger a_i -1\right) \\
		&+ \sum_{i,j\in Q\cup C} \hbar g_{ij} \left( a_i^\dagger a_j + a_i a_j^\dagger \right) + \hbar \frac{\Omega_0^x}{2} \left(  a_0^\dagger + a_0 \right), \label{eq:7}
\end{aligned}\end{equation}
which can be numerically solved to obtain an expected quantum operation.
We note that all quantum states involved in numerical simulations for quantum operations are dressed states~\cite{dressed_states}, which can be directly measured in real experiments.

In our simulations, the adjustable qubit frequencies are assumed to be in the range of $4.9$ to $5.1$ GHz, which are commonly used parameters in experiments. The coupler frequencies are aligned along the $ZZ$-free line, which are around $6.0$ GHz. Other parameters are assumed as follows. The anharmonicity of each qubit and coupler are $300$ MHz. The direct qubit-qubit coupling is $10$ MHz, the qubit-coupler coupling is $100$ MHz, the single-qubit driving strength is $25$ MHz, and the driving frequency matches the corresponding qubit frequency. Residual coupling between higher-order neighboring qubits due to parasitic capacitance is also taken into account~\cite{Crosstalk_zhaopeng,nnn_coupling}, which is $0.1$ MHz between $2$-hop neighboring qubits and $0.01$ MHz between $3$-hop neighboring qubits.

According to the definition, the quantum crosstalk errors from node $\mathcal{N}_i$ (or edge $\mathcal{E}_{ij}$) to node $\mathcal{N}_k$ are simulated.
When focusing on the single-qubit operations on qubit $Q_i$, and taking $R_X$ operations as an example, we assume that the quantum state of $Q_i$ is initialized to $\ket{\pm}_i=\left(\ket{0}_i+\ket{1}_i\right)/\sqrt{2}$, while the other qubits are initialized to their ground states. In our simulations, $20$ operation time durations are uniformly sampled in a range from $0$ to $2\pi/\Omega^x$, corresponding to rotating angles of $R_X$ operations uniformly sampled from $0$ to $2\pi$. Thus, the undesired excitations on qubit $Q_k$ across these $40$ scenarios are simulated and averaged to represent the crosstalk error from $Q_i$ to $Q_k$ when $R_X$ operations are implemented.
For two-qubit operations applied to the coupled-qubit pair $(Q_i,Q_j)$, the initial states are assumed to be $\ket{+}_i\ket{+}_j$ or $\ket{-}_i\ket{-}_j$, with the remaining qubits in their ground states. In our simulations, $20$ time durations ranging from $0$ to $2\pi/g_{XY}^{\rm eff}$ are also uniformly sampled. Similar to the approach taken for single-qubit operations, $40$ crosstalk excitations from coupled-qubit pair $(Q_i,Q_j)$ to $Q_k$ are simulated and averaged.
we ignore errors smaller than $1\times10^{-4}$, and cap errors larger than $1\times10^{-1}$ at $1\times10^{-1}$. This upper limit is introduced to avoid out-of-distribution errors. Data exceeding this threshold comprise less than $1\%$, as supported by Figs.~\ref{figs2}(a)-(i). Although introducing this upper limit does not affect the evaluator's MSE and $r^2$, it remains essential. The primary purpose of the evaluator is not only to fit the datasets, but also to guide the designer's training. During training, the designer may explore a wide range of data distributions. Without the upper limit, the designer might encounter data outside the distribution represented in the datasets. Since neural networks often exhibit significant inefficiency and uncertainty when inferring on out-of-distribution data, this would severely impact the designer's training. Therefore, it is necessary to strictly confine the evaluator's inputs and outputs to the dense region of the datasets' distribution.

The datasets for the evaluator are generated on the small graphs as shown in Fig.~\ref{figs1}(a), which represent the most common qubit coupling structures. Specifically, for each type of quantum operations, $25,600$ sets of node and edge frequencies are uniformly sampled on each $6$-qubit graph. And the corresponding quantum crosstalk errors are numerically simulated. In total, the datasets comprise $25,600 \times 5 {\rm (graphs)} = 128,000$ data points. Assuming that $100$ repeated measurements are required to obtain a single error value, the total number of measurements for one operation type would be $128,000 \times 2{\rm (initial\ states)} \times 20{\rm (time\ durations)} \times 100 = 512,000,000$. Given that a complete measurement cycle does not exceed $1\mu s$~\cite{readout_time}, it would take approximately $10$ minutes to generate the datasets for one quantum operation type in real experiments.
In the left of Fig.~\ref{figs1}(a), three small graphs are used to train the evaluator. Of these datasets, $80\%$ is allocated for training, $10\%$ for validation, and $10\%$ for testing. To ensure the evaluator's generalization ability to different graph structures, the two small graphs on the right side of Fig.~\ref{figs1}(a) are not used in training but only for testing purposes, with $10\%$ of their data forming a test set for new structures (NS). Additionally, to assess the evaluator's generalization to different graph scales, approximately $1,000$ data points for each type of quantum operations are generated on a graph with $30$ qubits coupled in a linear (one-dimensional) chain, serving as another test set (30Q).

\subsection{Simultaneous Operations}

Simultaneously applying quantum operations on different nodes or edges can improve the efficiency of quantum computing. Although the dataset for evaluator comprises only individual gates, the trained evaluator is also applicable for simultaneous operations.
One reason is that the undesired excitations caused by simultaneous operations are approximately the sum of these excitations induced by corresponding individual operations, owing to the linearity of Schr\"{o}dinger equations. Another reason is that the produced dataset includes all the possible cases from the node $\mathcal{N}_i$ (or edge $\mathcal{E}_{ij}$) to the node $\mathcal{N}_k$ with $k\neq i,j$, thereby all potential individual operation contributions to simultaneous gates are taken into account.

Numerical simulations on small graphs demonstrate the additivity of crosstalk errors.
Taking single-qubit operations in a $3$-qubit graph with the coupling configurations '$Q_0-Q_1-Q_2$' in which only connected two qubits have the direct coupling as examples, the average crosstalk excitations on qubit $Q_1$ are illustrated with dashed red lines in Figs.~\ref{figs7}(a) and (b) when $R_X$ (or $R_Y$) operations are simultaneously applied to qubit $Q_0$ and qubit $Q_2$.
The crosstalk errors, from the $R_X$ (or $R_Y$) operations separately applied to qubit $Q_0$ and qubit $Q_2$, are represented by solid blue lines. It is clear that the sums of the crosstalk errors resulted from the separate operations on two qubits approximately match to the errors resulted from the simultaneous operations on two qubits.
For two-qubit operations in a $5$-qubit graph with the coupling configuration `$Q_0-Q_1-Q_2-Q_3-Q_4$', simultaneous and separate $R_{XY}$ operations are applied on edge $\mathcal{E}_{01}$ and edge $\mathcal{E}_{34}$, and the crosstalk errors on qubit $Q_2$ are plotted in Fig.~\ref{figs7}(c).

We point out that it is worthless to apply a single-qubit operation and a two-qubit operation simultaneously, because single-qubit gates are typically much faster than two-qubit ones.
It is important to note that Figs.~\ref{figs7}(a)-(c) just use simple examples to show the additivity of the errors. In these examples, the position of two simultaneous operations are so close that the crosstalk errors resulted from the simultaneous operations are too large as shown in Fig.~\ref{figs7}(c). In practice, according to our simulations, the recommended minimum number of intermediate qubits for two simultaneous gates should be larger than two.

\section{Theory-Inspired Evaluator}\label{theory_evaluator}

In the main text, the MLPs-based evaluator is employed to estimate quantum crosstalk errors, achieving a good fit with $r^2$ exceeding $0.9$ across all test sets. Since the MLPs-based evaluator does not heavily rely on theoretical backgrounds, we also introduce a theory-inspired evaluator to further demonstrate the robustness of the designer's advantages. The theory-inspired evaluator contains only a few parameters, resulting in a slightly lower fitting accuracy compared to the MLPs-based evaluator in the main text. However, training the Theory-inspired evaluator requires significantly less data than the MLPs-based evaluator described in the main text, which enhances its experimental feasibility. And it still achieves an acceptable level of fitting performance.

\begin{table}[tbp]
	\caption{Results of theory-inspired evaluator, which is trained with smaller datasets}
	\label{tables1}
	\begin{ruledtabular}
		\begin{tabular}{lccc}
			\textbf{Dataset} & \textbf{$R_X$} & \textbf{$R_Y$} & \textbf{$R_{XY}$} \\
			\colrule
			training        & 0.037(0.91) & 0.037(0.92) & 0.042(0.92) \\
			validation      & 0.038(0.91) & 0.035(0.92) & 0.043(0.92) \\
			testing         & 0.040(0.90) & 0.044(0.89) & 0.041(0.92) \\
			testing (NS)    & 0.041(0.90) & 0.042(0.90) & 0.059(0.91) \\
			testing (30Q)   & 0.014(0.95) & 0.023(0.91) & 0.062(0.85) \\
		\end{tabular}
	\end{ruledtabular}
\end{table}

From a theoretical perspective, quantum crosstalk errors and optical mode interactions share a similar foundation in quantum optics, suggesting that the error model is expected to exhibit a Lorentzian line shape. Taking the crosstalk induced by single-qubit $R_X$ operations on qubit $Q_i$ to qubit $Q_k$ as an example, the error can be fitted using a parameterized Lorentzian line,
\begin{equation}
\hat{E}_{R_X,\mathcal{N}_i\to\mathcal{N}_k} = \frac{J_{d(i,k)}^2}{(\omega_i - \omega_k)^2 + J_{d(i,k)}^2},
\end{equation}
where $J_{d(i,k)}$ represents the effective coupling between the two qubits. In the theory-inspired evaluator, $J_{d(i,k)}$ is a learnable parameter, and $d(i,j)$ denotes the distance between node $\mathcal{N}_i$ and $\mathcal{N}_k$ in the graph. Due to the localized nature of quantum crosstalk, we only consider errors within the $4$-hop neighborhood.
The model for single-qubit $R_Y$ operations is consistent with that for $R_Y$ operations, while the $J_{d}$ parameters are trained based on the datasets of $R_Y$ operations.
For quantum crosstalk errors of two-qubit $R_{XY}$ operations, e.g., $ (Q_i, Q_j) \to Q_k $, the model incorporates Lorentzian line-shaped errors based on three frequency differences: $ \omega_{ij} - \omega_k $, $ \omega_i - \omega_k $, and $ \omega_j - \omega_k $. The resulting errors are weighted by their respective learnable weights and summed to represent the total quantum crosstalk error for $R_{XY}$ operations.
Similar to the MLPs-based evaluator in the main text, errors lower than $1\times10^{-4}$ are set to zero, while errors larger than $1\times10^{-1}$ are set to $1\times10^{-1}$. All the errors are trained and evaluated using base-$10$ logarithms.
The training and evaluation results for the theory-inspired evaluator are shown in Tab.~\ref{tables1}. The theory-inspired evaluator is trained using smaller datasets, with each training set containing only $30,720$ data points, corresponding to $61,440,000$ quantum measurements, which is $1/40$ of the training data required for the MLPs-based evaluator in the main text. The validation and testing of the theory-inspired evaluator still utilize the same datasets as the MLPs-based evaluator.
The $r^2$ values exceed $0.8$ across all datasets, demonstrating acceptable fitting performance.

\begin{figure}[htbp]
	\centering
	\includegraphics[width=1\linewidth]{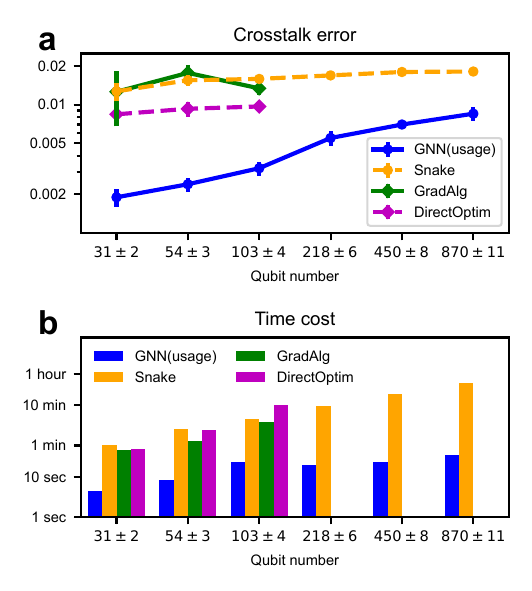}\\
	\caption{
		\textbf{The testing results based on the theory-inspired evaluator.} All the settings are consistent with that in the main text. According to the same theory-inspired evaluator, the GNNs-based designer achieves lower errors and faster speeds than Snake algorithm. This showcases the advantages of the designer are robust.
		}\label{figs8}
\end{figure}

Subsequently, the GNNs-based designer is trained and tested using the theory-inspired evaluator, following a procedure consistent with that described in the main text. The corresponding test results are shown in Fig.~\ref{figs8}. According to the theory-inspired evaluator, the GNNs-based designer achieves lower errors than the Snake algorithm across all SQEC scales. For instance, on SQECs with approximately $870$ qubits, the GNNs-based designer achieves an error of $0.0085$ within several seconds, whereas the Snake algorithm requires several minutes to reach an error of $0.0181$. These results demonstrate that the advantages of the GNNs-based designer are robust and broadly applicable.

\begin{figure*}[p]
	\centering
	\includegraphics[width=1\linewidth]{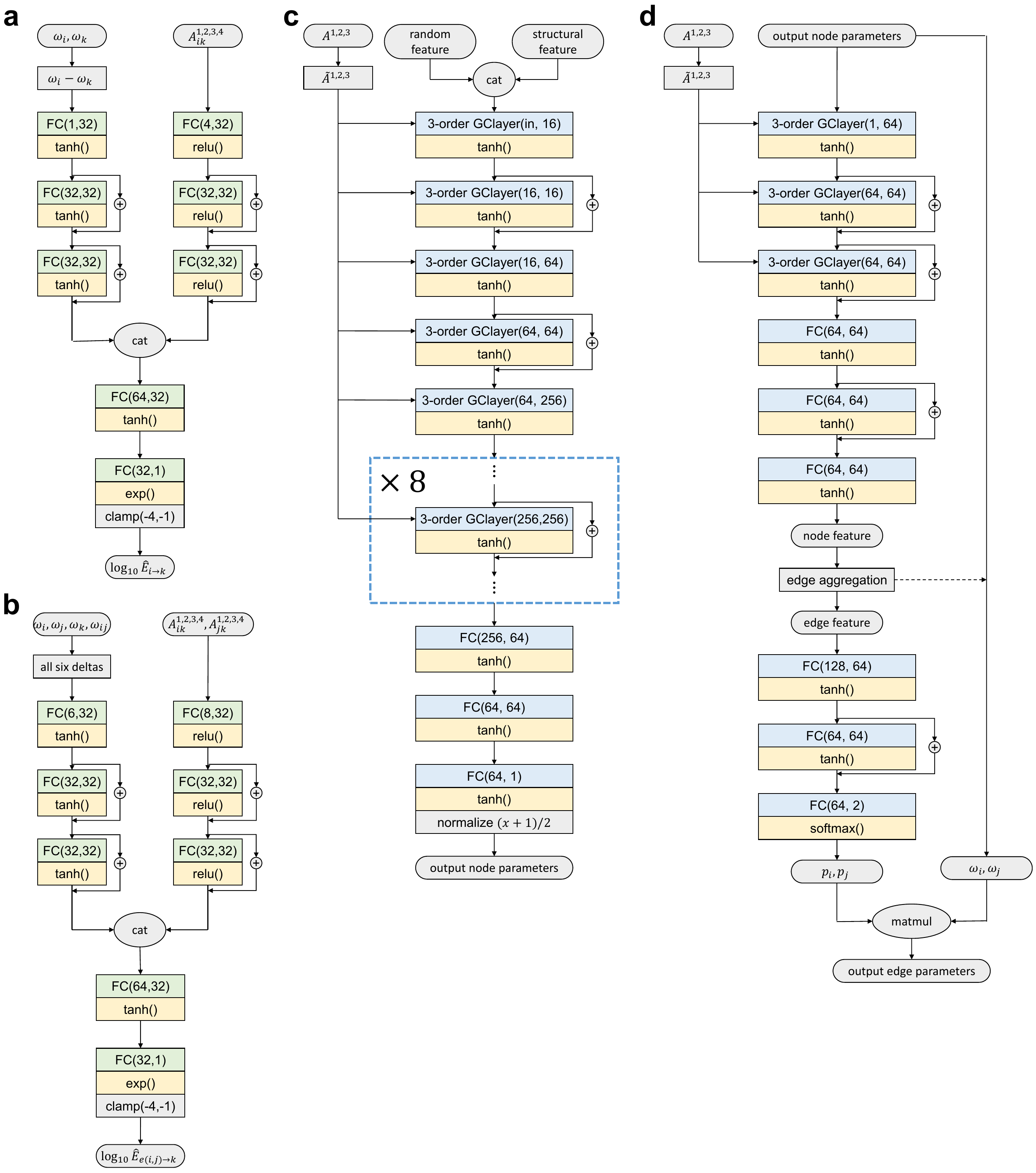}\\
	\caption{
			\textbf{The detailed architecture of the evaluator and the designer.} 
			\textbf{a.} The component of the evaluator for single-qubit operations.
			\textbf{b.} The component of the evaluator for two-qubit operations.
			\textbf{c.} The component of the designer for designing node parameters.
			\textbf{d.} The component of the designer for designing edge parameters, which takes the designed node parameters as input.
			\fbox{FC$(x,y)$} is a fully-connected layer with $x$ input features and $y$ output features.
			\fbox{3-order GCNlayer$(x,y)$} is a $3$-order graph convolutional layer with $x$ input features and $y$ output features.
			\fbox{tanh()}, \fbox{relu()}, \fbox{exp()}, \fbox{softmax()} are the non-linear activation functions.
			\fbox{+} represents adding two feature vectors.
			\fbox{matmul} refers to computing the matrix product of two vectors.
			\fbox{cat} means concatenating two feature vectors.
			\fbox{clamp(-4,-1)} means constraining the data within the range of $-4$ to $-1$.
			\fbox{all six deltas} in \textbf{b} are $\omega_i-\omega_j,\omega_i-\omega_k,\omega_j-\omega_k,\omega_{ij}-\omega_i,\omega_{ij}-\omega_j,\omega_{ij}-\omega_k$.
			\fbox{edge aggregation} in \textbf{d} refers to aggregating two node features into one edge feature for each edge within the graph.
		}\label{figs9}
\end{figure*}

\clearpage
\bibliography{ref}